\documentclass[aps,prd,floatfix,showpacs,preprintnumbers,twocolumn,superscriptaddress]{revtex4}
\usepackage{bm}
\usepackage{latexsym}
\usepackage{dcolumn}
\usepackage{amsfonts,amssymb,amsmath}
\usepackage{graphicx,epsfig}
\usepackage{psfrag}
\usepackage{multirow}
\def\tablefootnote#1{%
\hbox to
\textwidth{\hss\vbox{\hsize\captionwidth\footnotesize#1}\hss}}
\newcommand{\be}{\begin{eqnarray}}
\newcommand{\ee}{\end{eqnarray}}
\newcommand{\ud}{\mathrm{d}}

\newcommand{\lt}{\left(}
\newcommand{\rt}{\right)}
\newcommand{\lqu}{\left[}
\newcommand{\rqu}{\right]}

\newcommand{\dota}{\dot{a}}

\newcommand{\dotp}{\dot{\phi}}

\newcommand{\lag}{\mathcal{L}}

\newcommand{\lp}{\ell_{\rm p}}

\newcommand{\hub}{\mathcal{H}}

\newcommand{\beq}{\begin{equation}}
\newcommand{\eeq}{\end{equation}}
\newcommand{\pbg}{\phi_{_0}}
\newcommand{\dpbg}{\dot\phi_{_0}}
\newcommand{\ddpbg}{\ddot\phi_{_0}}
\newcommand{\Xbg}{X^{^{(0)}}\!}
\newcommand{\Pbg}{P^{^{(0)}}\!}
\newcommand{\PX}{P_{_X}\!}
\newcommand{\PXbg}{P_{_X}^{^{(0)}}\!}
\newcommand{\Pphi}{P_{_\phi}\!}
\newcommand{\Pphibg}{P_{_\phi}^{^{(0)}}\!}

\newcommand{\Pphiphibg}{P_{_{\phi\phi}}^{^{(0)}}\!}

\newcommand{\PXXbg}{P_{_{XX}}^{^{(0)}}\!}

\newcommand{\PXXXbg}{P_{_{XXX}}^{^{(0)}}\!}
\newcommand{\PXphi}{P_{_{X\phi}}\!}
\newcommand{\PXphibg}{P_{_{X\phi}}^{^{(0)}}\!}

\newcommand{\PXphiphibg}{P_{_{X\phi\phi}}^{^{(0)}}\!}

\newcommand{\PXXphibg}{P_{_{XX\phi}}^{^{(0)}}\!}

\def\a  {\alpha}
\def\b  {\beta}

\newcommand{\dophi}{\delta\phi}
\newcommand{\ddophi}{\delta\dot{\phi}}
\newcommand{\dddophi}{\delta\ddot{\phi}}
\newcommand{\di}{\delta^{^{(1)}}\!}
\newcommand{\dii}{\delta^{^{(2)}}\!}
\reversemarginpar
\begin{document}
\preprint{Preprint number: 0905.4184}
\title{A note on second-order perturbations of non-canonical scalar fields}
\author{Corrado~Appignani}
\email{appignani@bo.infn.it}
\affiliation{Dipartimento di Fisica, Universit\`a di
Bologna and I.N.F.N., Sezione di Bologna,
via Irnerio~46, 40126~Bologna, Italy}
\affiliation{Institute of Cosmology and Gravitation,
University of Portsmouth, Portsmouth PO1 3FX,
United Kingdom}
\author{Roberto~Casadio}
\email{casadio@bo.infn.it }
\affiliation{Dipartimento di Fisica, Universit\`a di
Bologna and I.N.F.N., Sezione di Bologna,
via Irnerio~46, 40126~Bologna, Italy}
\author{S.~Shankaranarayanan}
\email{shanki@iisertvm.ac.in}
\affiliation{Institute of Cosmology and Gravitation,
University of Portsmouth, Portsmouth PO1 3FX,
United Kingdom}
\affiliation{School of Physics, Indian Institute of Science Education 
and Research-Trivandrum, CET campus, Thiruvananthapuram 695 016, India}
\date{\today}
\begin{abstract}
We study second-order perturbations for a general non-canonical scalar
field, minimally coupled to gravity, on the unperturbed FRW
background, where metric fluctuations are neglected {\em a priori\/}.
By employing different approaches to cosmological perturbation theory,
we show that, even in this simplified set-up, the second-order
perturbations to the the stress tensor, the energy density and the
pressure display potential instabilities, which are not present at
linear order.
The conditions on the Lagrangian under which these instabilities
take place are provided.
We also discuss briefly the significance of our analysis in light of the
possible linearization instability of these fields about the
FRW~background.
\end{abstract}
\pacs{98.80.Cq,98.80.Jk,04.25.Nx}
\maketitle
\section{Introduction}
The physics of cosmological perturbations is a well-researched field
of
study~\cite{1963-Lifshitz.Khalatnikov-AdP,1966-Hawking-ApJ,1980-Bardeen-PRD,
1980-Lukash-ZhETF,1982-Chibisov.Mukhanov-MNRAS,1984-Kodama.Sasaki-PTPS,
1989-Ellis.Bruni-PRD,1992-Bruni.etal-CQG, 2005-L-V-PRD,
1992-Mukhanov.etal-PRep, 1994-Langlois-CQG,
1999-Ellis.Elst-NATO,2004-Bartolo.etal-PRep,2008-Tsagas.etal-PRep,2008-Malik.Wands-Arx,
2008-Malik.Matravers-CQG}.
At linear order, the (quantized) cosmological perturbation theory has been
the primary tool to investigate the behavior of fluctuations during inflation.
For instance, during the slow-roll phase, the linear perturbation theory
predicts an approximately scale-free spectrum of density fluctuations,
which is consistent with Cosmic Microwave Background (CMB)
measurements~\cite{2009-Komatsu.etal-ApJS}.
\par
During the last few years, considerable amount of attention has been
devoted to examine the effect of higher-order corrections to the
linearized Einstein's equations.
There are three main reasons which have led to such an enormous interest.
First, the study of higher-order perturbations is imperative to quantify the primordial
non-Gaussanity of the
CMB~\cite{1997-Bruni.etal-CQG,2003-Maldacena-JHEP,
2004-Bartolo.etal-PRep,2005-Seery.Lidsey-JCAP}, which will be
confronted with the data collected by the PLANCK
mission~\cite{2005-Planck-ScientificProgrammeof}.
Second, within the linear theory, it is not possible to determine when
the perturbations become large and non-linearities should be taken into account.
For instance, gravitational waves in Minkowski space-time can have
arbitrary amplitudes in the linear perturbation theory.
The only way to understand the extent of the backreaction of the perturbations on
the Friedman-Robertson-Walker (FRW)~background is therefore to
consider at least the second
order~\cite{1997-Abramo.etal-PRD,1999-Ellis.Elst-NATO,2008-Buchert-GRG}.
Third, higher-order corrections may help to explain the dark energy.
For example, there has been a radical proposal to abandon the
Copernican principle and, instead, suppose that we are near the center
of a large, non-linearly under-dense, nearly spherical void surrounded
by a flat, matter dominated space-time~(For recent reviews, see
Refs.~\cite{2007-Wiltshire-Talk,2008-Enqvist-GRG}).
\par
There are four different approaches in the literature to study
cosmological perturbations:
\begin{description}
\item[{\tt 1)}]~solving Einstein's equations
order-by-order~\cite{1963-Lifshitz.Khalatnikov-AdP};
\item[{\tt 2)}]~the covariant approach based on a general frame vector
$u^{\alpha}$~\cite{1966-Hawking-ApJ,1989-Ellis.Bruni-PRD,1992-Bruni.etal-CQG,
2005-L-V-PRD, 1999-Ellis.Elst-NATO};
\item[{\tt 3)}]~the Arnowitt-Deser-Misner (ADM) approach based on the normal frame vector
$n^{\alpha}$~\cite{1980-Bardeen-PRD,1984-Kodama.Sasaki-PTPS,
1992-Mukhanov.etal-PRep,1994-Langlois-CQG};
\item[{\tt 4)}]~the reduced action
approach~\cite{1980-Lukash-ZhETF,1982-Chibisov.Mukhanov-MNRAS}.
\end{description}
In the case of linear perturbations, it has been shown that all of
these four approaches lead to identical equations of motion.
However, to our knowledge, a complete analysis has not been done in the literature
for higher-order perturbations (for an earlier study on the differences between
the approaches {\tt 1)}~and {\tt 3)}~above, see Ref.~\cite{MSA}).
This may be attributed to the following reasons:
(a)~unlike the linear order, the scalar, vector and tensor perturbations
do not decouple and can not be treated independently and
(b)~although certain physical quantities are gauge-invariant
(like tensor metric perturbations) at first order, they may become
gauge-dependent at second order~\cite{1997-Bruni.etal-CQG}.
Hence, to obtain gauge-invariant definitions of physically relevant quantities
at second (or higher) order is far more complicated~\cite{2008-Malik.Wands-Arx}.
This leads to certain observables having completely different values in different
frames.
\par
In this note, to illustrate the problems that may occur at higher-order,
and not to get bogged-down with the gauge issues, we consider a
simple situation:
we freeze all the metric perturbations and focus on the perturbations of a
minimally-coupled, generalized scalar field $\phi$, whose Lagrangian
density is given by~\cite{1999-Armendariz-Picon.etal-PLB}
\beq
\label{eq:genscal}
\lag = P(X, \phi)
\ ,
\qquad
\mbox{where}
\qquad
2\,X = \nabla^{\alpha} \phi\,\nabla_{\alpha} \phi
\ .
\eeq
More precisely, we will only consider linear perturbations of
the scalar field,
\beq
\label{eq:def-Perphi}
\phi(t, {\bf x})=\pbg(t) + \dophi(t, {\bf x})
\ ,
\eeq
about the four-dimensional FRW~background,
\beq
\label{eq:FRW}
\ud s^2 = N^2\,\ud t^2-\gamma_{ij}\,\ud x^i\,\ud x^j
=\ud t^2-a^2(t)\,\ud{\bf x}^2
\ ,
\eeq
while expanding all the dependent quantities, like $X$ and the stress
tensor, up to second order~\footnote{Our analysis is similar in spirit
to Gruzinov's calculation~\cite{2005-Gruzinov-PRD}.}.
We again wish to emphasize that our aim is to highlight ambiguities
which occur at second order, and not to solve the second-order Einstein
equations.
In particular, we shall obtain second-order quantities of physical
relevance from the scalar field Lagrangian~\eqref{eq:genscal}, in this
simplified set-up, by employing different approaches and highlight the
main differences.
\par
To exemplify such differences, it will appear convenient to compare the
ratio
\be
\label{eq:cs-def}
c_{\rm s}^2 = \frac{\mbox{coefficients of~} (\dophi_{,i}/a)^2}
{\mbox{coefficients of~}
\ddophi^2}
\ ,
\ee
in the components of the stress~tensor and related quantities [see
Eqs.~\eqref{eq:d2Tmunu},~\eqref{eq:d2rp} and~\eqref{eq:shao} below].
Since $c_{\rm s}^2$ is dimensionally the square of a speed, we will refer
to this ratio as the ``speed of propagation''.
We will also discuss the relation between these $c_{\rm s}^2$ and the
square of the phase velocity $c^2_{\delta \phi}$ derived from the wave
equation for the perturbation $\delta \phi$ [for small perturbations,
Eq.~\eqref{eq:cs-def} coincides with the standard definition in the theory of
elasticity~\cite{1970-Landau.Lifshitz-Elasticity}]
and of the adiabatic speed of sound
$c^2_\textsc{ad} = \partial p / \partial \rho$, with $p$ the pressure
and $\rho$ the energy density.
We will then show that the second-order stress~tensor $\dii T_{00}$ and
the second-order canonical Hamiltonian obtained from the reduced action,
which coincide for a canonical scalar field, are in general different.
\par
Among our results, the one which seems of major physical concern will be the
emergence of instabilities in the quantities analysed.
For example, we shall show that the second-order energy density $\dii \rho$,
as defined in the covariant approach, takes a negative contribution from the spatial
derivative term, which is also present for the canonical scalar field.
Similar terms, which lead to large negative contributions for short wave-lengths,
generically appear in the components of the second-order stress~tensor
and are signalled by imaginary speeds of propagation, that is $c_{\rm s}^2<0$.
This finding poses a serious question about the applicability of the perturbative
approach (at least) for non-linear scalar field Lagrangians of the
form~\eqref{eq:genscal}.
\par
Our metric signature is $(+,-,-,-)$ and lower case Greek (Latin)
indices refer to four (three) dimensions.
Time derivatives are denoted by a dot and, for any variable $G(X,\phi)$,
its background unperturbed value is denoted by $G^{^{(0)}}$,
that is $G^{^{(0)}}\equiv G(X(\pbg),\pbg)$.
Also, for any scalar function $f$, we will be using the notations
$f^2_{,i} \equiv \sum_{i=1}^{3} (f_{,i})^2$ and
$f_{,ii} \equiv \sum_{i=1}^{3} f_{,ii}$
throughout the paper.
\section{Equation of motion and stress tensor}
The equation of motion can be easily derived from the action principle for
the general Lagrangian~\eqref{eq:genscal},
\be
\label{eq:eom}
\PX\,\nabla_{\mu}\nabla^{\mu}\phi
+\left(\nabla_{\mu}\PX\right)\nabla^{\mu}\phi
-\Pphi = 0
\ ,
\ee
and the corresponding stress tensor is given by
\be
\label{eq:Tmunu}
T_{\mu\nu}=\PX\, \nabla_{\mu} \phi\, \nabla_{\nu} \phi - g_{\mu\nu}\,P
\ ,
\ee
where $\PX\equiv \partial P/\partial X$,  $\Pphi\equiv\partial P/\partial \phi$
and so on.
Eqs.~\eqref{eq:eom} and \eqref{eq:Tmunu} are manifestly covariant
and hold in arbitrary space-times.
We then set out to perturb both expressions up to second order, neglecting
the metric perturbations.
For instance, the most general perturbation of the stress tensor to all orders
can be written as
\be
T_{\mu\nu} &\!\!=\!\!&
\PX\!\left(\Xbg+\Delta X, \pbg+ \Delta \phi \right)
\nonumber
\\
&&
\times
\nabla_{\mu}\!\left(\pbg+ \Delta \phi\right)
\nabla_{\nu}\!\left(\pbg+ \Delta \phi\right)
\nonumber
\\
&&
-\left(g^{^{(0)}}_{\mu\nu} + \Delta g_{\mu\nu} \right)
P\!\left(\Xbg+ \Delta X,\pbg+\Delta\phi\right)
\ ,
\label{Tallord}
\ee
where $\Delta \phi$, $\Delta X$ and $\Delta g_{\mu\nu}$ represent
perturbations to all orders.
We then freeze the metric perturbations by setting
$\Delta g_{\mu\nu} = 0$ and expand the other terms up to second order
using the form in Eq.~\eqref{eq:def-Perphi}.
\subsection{Perturbed equation of motion}
We start from the equation of motion.
The evolution for the background scalar field $\pbg$ is determined
by
\be
\lt \PXbg  + \PXXbg\,\dpbg^2 \rt \ddpbg
+ \PXphibg\, \dpbg^2 + 3\,\frac{\dota}{a}\,\PXbg\, \dpbg - \Pphibg
=0
\ ,
\ee
while the dynamics of the perturbation $\delta \phi$ is governed by
\begin{subequations}
\be
\lt \PXbg  + \PXXbg\, \dpbg^2 \rt \dddophi
- \frac{\PXbg}{a^2}\,\dophi_{,ii}
+C\,\ddophi+D\,\dophi=0
\ ,
\label{1steom}
\ee
with
\be
C&\!\!=\!\!&
3\, \frac{\dota}{a} \lt \PXbg +  \PXXbg\,\dpbg^2 \rt
+ \lt 3\, \PXXbg\,\ddpbg + \PXphi \rt\dpbg
+ \PXXbg
\nonumber
\\
&&
+ \lt \PXXphibg +  \PXXXbg\,\ddpbg \rt  \dpbg^3
\label{coeC}
\\
D&\!\!=\!\!&
3\,\frac{\dota}{a} \, \PXphibg\,\dpbg +
\lt \PXphiphibg +  \PXXphibg\,\ddpbg \rt \dpbg^2
\nonumber
\\
&&
+  \PXphi\,\ddpbg- \Pphiphibg
\ .
\label{coeD}
\ee
\end{subequations}
This is a wave equation from which we can immediately read off
the speed of propagation of the field $\dophi$, namely
\beq
c^2_{\dophi} = \frac{\PXbg}{\PXbg  +  \PXXbg\,\dpbg^2}
\ ,
\label{1stordersos}
\eeq
which matches the speed of sound used in most of
the literature (see, e.g.~Ref.~\cite{1999-Garriga.Mukhanov-PLB}).
When $c_{\dophi}^2<0$, we therefore expect that the dynamics becomes
unstable (see, e.g.~Ref.~\cite{1999-Armendariz-Picon.etal-PLB}).
The nature of this instability is easily understood using
an analogy with classical mechanics:
when $c_{\dophi}^2$ is negative, the system resembles
an inverted harmonic oscillator and, no matter how small the amount of
perturbation $\delta \phi$, it will rapidly run away from the
background solution $\pbg$ and from the perturbative regime.
\par
Note also that the term $C\,\ddophi$
in Eq.~\eqref{1steom} would
make the frequency $\omega_k$ of Fourier modes
$\dophi_k$ complex, and the perturbation $\dophi_k$
would thus decay or grow exponentially in the proper time $t$.
However, terms containing $\ddophi$ can be eliminated by
rescaling
\be
\dophi(t,{\bf x})\to  r(t) \,\chi(t,{\bf x})
\ ,
\label{chi}
\ee
where $r$ is a suitable function of the background quantities,
and this kind of behaviour can thus be studied within the perturbative
approach.
But this procedure does not change the speed of
propagation~\eqref{1stordersos} [nor those defined according
to Eq.~\eqref{eq:cs-def}, see Appendix~\ref{rescale})]
and cannot remove the associated instabilities.
In the following, we shall therefore focus only on the instabilities
signalled by imaginary speeds of propagation.
\subsection{Perturbed stress tensor}
We now turn our attention to the stress tensor and expand it to
second order,
\be
T_{\mu\nu} = T_{\mu\nu}^{^{(0)}} + \di T_{\mu\nu} + \dii T_{\mu\nu}
\ ,
\ee
with
\be
\label{eq:0Tmunu}
T_{\mu\nu}^{^{(0)}}
=\PXbg\,\dpbg^2\, \delta^0_{\,\mu}\, \delta^0_{\,\nu} - g_{\mu\nu}\,\Pbg
\ .
\ee
Linear perturbations are then given by
\begin{subequations}
\label{eq:d1Tmunu}
\be
\label{eq:d1T00}
\!\!\!\!\!\!\!\!\!\!\!\!
\di T_{00}
&\!\!=\!\!&
\left[\PXbg+\PXXbg\,\dpbg^2\right]\!
\dpbg\,\ddophi
-\left[\Pphibg-\PXphibg\,\dpbg^2\right]
\!\dophi
\\
\label{eq:d1T0i}
\!\!\!\!\!\!\!\!\!\!\!\!
\di T_{0i}
&\!\!=\!\!&
\PXbg\,\dpbg\,\dophi_{,i}
\\
\label{eq:d1Tij}
\!\!\!\!\!\!\!\!\!\!\!\!
\di T_{ij}
&\!\!=\!\!&
a^2 \left(\PXbg\,\dpbg\,\ddophi +  \Pphibg\, \dophi\right) \delta_{ij}
\ ,
\ee
\end{subequations}
and second-order perturbations by
\begin{subequations}
\label{eq:d2Tmunu}
\begin{eqnarray}
\dii T_{00}
&\!\!=\!\!&
\left(\PXbg + 4\,\PXXbg\,\dpbg^2+\PXXXbg\,\dpbg^4\right)
\frac{\ddophi^2}{2}
\nonumber
\\
&&
+ \left(\PXbg-\PXXbg\,\dpbg^2\right)\!
\frac{\dophi_{,i}^2}{2\,a^2}
- \left(\Pphiphibg-\PXphiphibg\,\dpbg^2\right)
\!\frac{\dophi^2}{2}
\nonumber
\\
\label{eq:d2T00}
&&
+
\left(\PXphibg+\PXXphibg\,\dpbg^2\right) \dpbg\, \dophi\,\ddophi
\\[2mm]
\dii T_{0i}
&\!\!=\!\!&
\left(\PXbg + \PXXbg\,\dpbg^2\right) \ddophi\,\dophi_{,i}
+\PXphibg\,\dpbg\,\dophi\,\dophi_{,i}
\label{eq:d2T0i}
\\[2mm]
\dii T_{ij}
&\!\!=\!\!&
\PXbg \lt 1-\frac{\delta_{ij}}{2} \rt \dophi_{,i}\,\dophi_{,j}
\nonumber
\\
&&
+\,\delta_{ij} \frac{a^2}{2}
\bigg[ \left(\PXbg + \PXXbg\, \dpbg^2 \right) \ddophi^2
\nonumber
\\
&&
+ \,2\,\PXphibg\,\dpbg\,\dophi\,\ddophi
+ \,\Pphiphibg\,\dophi^2 \big] \,.
\label{eq:d2Tij}
\end{eqnarray}
\end{subequations}
\par
We would now like to stress the following points regarding the perturbed
stress tensor:
\\
{\sl i)}~the components $\di T_{\mu\nu}$ are identical to the
expressions given in Ref.~\cite{1999-Garriga.Mukhanov-PLB} for the
case when the metric perturbations are frozen.
\\
{\sl ii)}~for an arbitrary scalar field Lagrangian, $\dii T_{00}$ may
represent an unstable perturbation.
Indeed, by expanding the perturbations in Fourier modes
(so that $\dophi_{,i}^2\sim k^2\,\dophi_k^2$),
one finds that the ratio between the coefficients
of $\dophi_{,i}^2$ and $\ddophi^2$ can in general be negative
(and become large for large $k$ and/or small $a$).
The origin of this instability is similar to the one
we already discussed with regard to the speed of
the perturbation obtained from the equation of motion.
We will say more on this point later, by considering specific
non-canonical Lagrangians, and only remark here that for the canonical
scalar field, i.e.~for
\be
P = X - V(\phi)
\ ,
\label{canphi}
\ee
this problem is not present, since
\be
\dii T_{00}^{^{\rm (KG)}}
= \frac{\ddophi^2}{2} + \frac{\dophi_{,i}^2}{2\,a^2} +
\frac{V_{\phi\phi}}{2} \,\dophi^2
\ .
\ee
\\
{\sl iii)}~by the same token, we observe that $\dii T_{ii}$ is
potentially unstable.
In this case it is the ratio between $\PXbg$ and $\PXbg+\PXXbg\,\dpbg^2$
which determines the stability of the system.
If this ratio is negative, the second-order pressure perturbations are unstable.
\\
{\sl iv)}~only under very special conditions, most notably for the
canonical scalar field, the {\it effective speed of propagation\/} of
$\dii T_{00}$ and $\dii T_{ii}$ are the same as that in
Eq.~\eqref{1stordersos} and equal to unity.
Using the definition~\eqref{eq:cs-def}, we can define
a speed related with the propagation of energy density perturbations
in the background frame from $\dii T_{00}$, that is
\begin{subequations}
\be
c_{0}^2=
\strut\displaystyle\frac{\PXbg - \PXXbg\, \dpbg^2}
{\PXbg +4\,\PXXbg\,\dpbg^2+ \PXXXbg\, \dpbg^4}
\ ,
\label{c1}
\ee
and a speed for momentum perturbations from $\dii T_{ii}$,
\be
c_{\parallel}^2 = c^2_{\delta \phi}
\ ,
\label{c2}
\ee
\end{subequations}
which may be different for non-canonical scalar fields (due to the
non-linearity of the dynamics).
One then immediately notes that these velocities become imaginary
right in correspondence with the instabilities mentioned previously
in {\sl iii)} and {\sl iv)}.
Finally, it is important to note that it is $c_{\parallel}$
which equals the speed of perturbations for non-canonical scalar fields
given in the literature~\cite{1999-Garriga.Mukhanov-PLB}, and
we will elaborate about the importance of this result when we discuss
the symmetry reduced action.
\section{Covariant approach}
The covariant approach~\cite{1966-Hawking-ApJ} relies upon the
introduction of a family of observers travelling with a time-like
four-velocity $u^{\mu}$.
By means of $u^{\mu}$, all the (geometrical) physical objects
and operators are decomposed into invariant parts:
the scalars along $u^{\mu}$ and scalars, three-vectors, and projected,
symmetric and trace-free tensors orthogonal to $u^{\mu}$.
Einstein's equations are then supplemented by the Ricci identities
for $u^{\mu}$ and the Bianchi identities, forming a complete set of first-order
differential equations (details can be found in
Refs.~\cite{1989-Ellis.Bruni-PRD,1992-Bruni.etal-CQG,1999-Ellis.Elst-NATO,
2008-Tsagas.etal-PRep}).
\par
The stress~tensor for a general scalar field~\eqref{eq:Tmunu}
then takes the perfect fluid form
\be
\label{eq:Tmunu-fluid}
T_{\mu\nu} = (\rho + p)\, u_{\mu}\, u_{\nu} - p\, g_{\mu\nu}
\ ,
\ee
if the time-like unit vector $u_{\mu}$ is chosen
as~\cite{1988-Madsen-CQG,1991-Ellis.Madsen-CQG}
\be
\label{eq:umu}
u_{\mu} = \frac{\nabla_{\mu} \phi}{\sqrt{2\,X}}
\ .
\ee
In the above, $\rho$ and $p$ are, respectively, the energy density and
pressure along the fluid flow and are given by
\be
\label{eq:def-rhop}
\rho = T_{\mu \nu}\, u^{\mu}\, u^{\nu}
\ ,
\quad
{\rm and}
\quad
p = -\frac{1}{3}\,T_{\mu\nu}\, h^{\mu\sigma}\, h^{\nu}_{\ \sigma}
\ ,
\ee
where $h_{\mu\nu} = g_{\mu\nu} - u_{\mu}\, u_{\nu}$ is the metric on a slice of fixed
observer's time.
\par
On expanding $u_{\mu}$, $T_{\mu\nu}$ and $h_{\mu\nu}$ up to second order,
we obtain
\begin{subequations}
\be
&
\rho^{^{(0)}} = T_{00}^{^{(0)}}
\ ,
\quad
\di\rho = \di T_{00}
&
\nonumber
\\
&
\dii\rho =
\dii T_{00}
- \strut\displaystyle\frac{1}{\PXbg}
\left(\strut\displaystyle\frac{\di T_{0i}}{a\, \dpbg}\right)^2
&
\label{eq:delta2rho}
\ee
and
\be
&
p^{^{(0)}} = P^{^{(0)}}
\ ,
\quad
\di p =  \strut\displaystyle - \frac{\di T^i_{\ i} }{3}
&
\nonumber
\\
&
\dii p = \strut\displaystyle\frac{1}{3\,a^2} \left[\dii T_{ij} \, \delta^{ij}
- \frac{1}{\PXbg} \left(\frac{\di T_{0i}}{\dpbg}\right)^2 \right]
\ .
&
\label{eq:delta2p}
\ee
\end{subequations}
Substituting~\eqref{eq:d1Tmunu} and~\eqref{eq:d2Tmunu} into the above
equations yields the second-order corrections
\begin{subequations}
\label{eq:d2rp}
\be
\dii \rho
&\!\!=\!\!&
\left(\PXbg + 4\,\PXXbg\,\dpbg^2 + \PXXXbg\,\dpbg^4 \right)
\frac{\ddophi^2}{2}
\nonumber
\\
&&
-\left(\PXbg + \PXXbg\,\dpbg^2\right) \frac{\dophi_{,i}^2}{2\,a^2}
- \left( \Pphiphibg-\PXphiphibg\,\dpbg^2\right) \frac{\dophi^2}{2}
\nonumber
\\
\label{eq:d2rho}
&&
+\left(\PXphibg +\PXXphibg\, \dpbg^2\right) \dpbg\,\dophi\,\ddophi
\\
\dii p
&\!\!=\!\!&
\frac{1}{2} \left(\PXbg + \PXXbg\, \dpbg^2 \right) \ddophi^2
+\PXphibg\,\dpbg\,\dophi\,\ddophi
\nonumber
\\
\label{eq:d2p}
&&
-\frac{\PXbg}{2\,a^2}\,\dophi_{,i}^2
+\frac{\Pphiphibg}{2}\,\dophi^2
\ .
\ee
\end{subequations}
We would then like to stress the following points regarding the perturbed energy
density and pressure:
\\
{\sl i)}~Eqs.~\eqref{eq:delta2rho} and~\eqref{eq:delta2p} show that,
up to linear order, the energy density and pressure measured in the
fluid frame are identical to the same quantities evaluated along the
cosmic time.
For example, $\rho=T_{00}\,u^{0}\,u^{0}$ up to linear order.
At second (and higher) order, however, the energy densities
measured by these two different observers are no more equal,
suggesting that general relativistic effects appear from the second
order on.
The results obtained here are in fact similar to the corrections
derived in the parameterized post-Newtonian
formulation~\footnote{See, for instance, Sec.~(39.7) in
Ref.~\cite{1973-Misner.etal-Gravitation}.}.
\\
{\sl ii)}~in the fluid frame, the energy density exhibits the same
kind of instability we found in the previous section for $\dii T_{00}$,
but this time the problem is present also for the canonical
scalar field, because of the negative contribution coming from $\di T_{0i}$.
In fact, substituting the Lagrangian~\eqref{canphi} in
Eq.~\eqref{eq:d2rho} yields
\be
\dii\rho^{^{\rm (KG)}}=
\frac{\ddophi^2}{2}
- \frac{\dophi_{,i}^2}{2\,a^2}
+ \frac{V_{\phi\phi}}{2} \,\dophi^2
\ee
so that, using again the analogy with classical mechanics, the
perturbations turn out to be unstable because of the negative sign of
the second term in the right hand side (which dominates over the potential
term for small $a$ and, in the Fourier domain, for large wavenumber $k$).
Although the results in the two frames (fluid and background) are related
by a Lorentz transformation, the authors could not find a discussion of
such an instability in standard textbooks~\footnote{For example, in
Ref.~\cite{1990-Kolb.Turner-Earlyuniverse} (Sec.~8.3, page~276),
$\dii\rho$ is claimed to be positive, but no proof is given from first
principles.
We thank D.~Wands for pointing this out.}.
\\
{\sl iii)}~only under special conditions, the {\it effective speed of
propagation\/} of the energy density and pressure perturbations are
equal (as in the previous Section, this occurs for the canonical
scalar field).
Using the definition~\eqref{eq:cs-def}, the speed of propagation
for density perturbations in the fluid frame turns out to be given by
\begin{subequations}
\be
c_{\rho}^2
=
- \frac{\PXbg + \PXXbg\,\dpbg^2}
{\PXbg +4\,\PXXbg\,\dpbg^2 +  \PXXXbg\, \dpbg^4}
\ ,
\label{crho}
\ee
and the velocity of pressure perturbations by
\be
c_{p}^2=c_{\parallel}^2=c_{\delta\phi}^2
\ ,
\ee
\end{subequations}
from Eq.~\eqref{c2}, and they are obviously different in general.
For completeness, we recall that the {\em adiabatic speed of sound\/}
(see, e.g.~Ref.~\cite{ch-malik}) is given by
\be
c^2_\textsc{ad} = \left.\frac{\partial p}{\partial \rho}\right|_{S} =
\frac{\PXbg\,\ddpbg + \Pphibg}
{\PXbg\,\ddpbg - \Pphibg + \PXXbg\,\dpbg^2\,\ddpbg + \PXphibg\,\dpbg^2}
\ ,
\ee
and differs from the other expressions shown so far, and in particular
from~\cite{ch-malik}
\be
c_{\rm s}^2=\frac{p_{_X}}{\rho_{_X}}=c_{\delta\phi}^2
\ .
\ee
\section{ADM~approach}
In the ADM~formulation~\cite{2008-Arnowitt.etal-GRG},
the Einstein-Hilbert action with matter can be written as
\be
\label{eq:ADM-action}
S = \int \ud t\,\ud^3x \left(\pi^0\,\dot{N} + \pi^i\,\dot{N}_i - N\,H - N_i\,H^i\right)
\ ,
\ee
where $N$ and $N^{i}$ are the lapse and shift functions, respectively,
and $\pi^0$ and $\pi^i$ their conjugate momenta.
The super-Hamiltonian and super-momenta are given by
\beq
\label{eq:def-superH}
H = - \frac{\partial S}{\partial N}
\ ,
\quad
H^i= - \frac{\partial S}{\partial N_i}
\ .
\eeq
\par
Expanding the Lagrangian~\eqref{eq:genscal} to second order about the
FRW~background~\eqref{eq:FRW} (with $N_i=0$) leads to
\be
P(X, \phi)
&\!\!\simeq\!\!&
\Pbg
+
\left({\PXbg}\,\frac{\dpbg\,\ddophi}{N^2}+\Pphibg\,\dophi\right)
\nonumber
\\
&&
+
\left(\PXbg + \PXXbg\,\frac{\dpbg^2}{N^2} \right)
\frac{\ddophi^2}{2\,N^2}
+ \frac{\Pphiphibg}{2}\,\dophi^2
\nonumber
\\
&&
-\frac{\PXbg}{2\,a^2}\,\dophi_{,i}^2
+\PXphibg\,\frac{\dpbg\,\ddophi}{N^2}\,\dophi
\ .
\label{eq:action2order}
\end{eqnarray}
Substituting the perturbed action in Eq.~\eqref{eq:def-superH}
(and setting $N = 1$) then leads to
\be
H =H^{^{(0)}} + \di H + \dii H
\ ,
\ee
where
\begin{subequations}
\label{eq:shao}
\begin{eqnarray}
\label{eq:0SH}
H^{^{(0)}}
&\!\!=\!\!&
\PXbg\,\dpbg^2 -  \Pbg
\\
\label{eq:d1SH}
\di H
&\!\!=\!\!&
\left(\PXbg+\PXXbg\,\dpbg^2 \right)\,\dpbg\,\ddophi
\nonumber
\\
&&
-\left(\Pphibg-\PXphibg\,\dpbg^2\right)\dophi
\\
\label{eq:d2SH}
\dii H
&\!\!=\!\!&
\left(\PXbg +4\,\PXXbg\,\dpbg^2 + \PXXXbg\,\dpbg^4 \right)\frac{\ddophi^2}{2}
\nonumber
\\
&&
+\left(\PXbg - \PXXbg\,\dpbg^2\right)\frac{\dophi_{,i}^2}{2\,a^2}
- \left(\Pphiphibg-\PXphiphibg\,\dpbg^2 \right) \frac{\dophi^2}{2}
\nonumber
\\
&&
+\left(\PXphibg+\PXXphibg\,\dpbg^2\right) \dpbg \, \dophi\,\ddophi
\ .
\label{sh2nd}
\end{eqnarray}
\end{subequations}
Looking at \eqref{sh2nd} we note the following:
\\
{\sl i)}~at all orders, the super-Hamiltonian is identical to the
$00$-component of the stress tensor given in
Eqs.~\eqref{eq:0Tmunu},~\eqref{eq:d1T00} and~\eqref{eq:d2T00}.
Although this might seem obvious, we would like to point out that the
two quantities are derived in different ways:
the stress tensor is obtained from the complete matter action,
whereas the super-Hamiltonian is obtained from the symmetry reduced
(and perturbed) action for the matter alone.
We will discuss more on this aspect below.
\\
{\sl ii)}~like in the approach of the perturbed stress tensor,
the {\it effective speed of propagation\/} of $\dii H$ defined in
Eq.~\eqref{eq:cs-def} is given by $c_0$ from Eq.~\eqref{c1} and
is identical to the speed of sound used in the
literature~\cite{1999-Garriga.Mukhanov-PLB}
only under special conditions [satisfied by the canonical scalar
field~\eqref{canphi}].
\section{Symmetry-reduced action}
Following the seminal works of Lukash~\cite{1980-Lukash-ZhETF}, and
Chibisov and Mukhanov~\cite{1982-Chibisov.Mukhanov-MNRAS},
this procedure has been extensively used in quantifying primordial
perturbations and their non-Gaussianity from
inflation~\cite{2003-Maldacena-JHEP,2005-Seery.Lidsey-JCAP}.
The basic idea is to perturb the action about the FRW~background,
up to second (or higher) order, and reduce it so that the perturbations are
described in terms of a single gauge-invariant variable,
which will depend on the metric and matter content.
\par
Here, our aim is to obtain the canonical Hamiltonian $\mathcal{H}$
corresponding to the perturbations of the generalized scalar field
and compare with the quantities previously derived in the other approaches.
Using the perturbed action up to second order from Eq.~\eqref{eq:action2order},
and decomposing the modes in the Fourier domain, gives the following
second-order action for the matter perturbations $\dophi_k$:
\be
\dii S
&\!\!=\!\!&
\int \ud t\, \frac{a^3}{2} \left[\left(\PXbg + \PXXbg \dpbg^2 \right) {\ddophi}_k^2
+
2\,\PXphibg\,\dpbg\,\dophi_k\,{\ddophi}_k
\right.
\nonumber
\\
&&
\left.
\phantom{\int \ud t\, \frac{a^3}{2}}
+\left(\Pphiphibg- \frac{k^2}{a^2}\,\PXbg\right){\dophi}_k^2
\right]
\ .
\label{eq:2orderaction}
\ee
Defining the canonical momentum conjugate to $\dophi_k$ as
\be
P_k = \frac{\partial \delta^{(2)} S}{\partial \ddophi_k}
\ ,
\ee
the canonical Hamiltonian corresponding to the perturbed
action~(\ref{eq:2orderaction}) reads
\beq
\dii \mathcal{H} =
\frac{a^3}{2}\!
\left[\!
\left(\PXbg +  \PXXbg\,\dpbg^2 \right)\! \ddophi_k^2
\!+\!\left(\frac{k^2}{a^2}\,\PXbg - \Pphiphibg \right)\! \dophi_k^2
\right]
.
\label{eq:d2canH}
\eeq
The point that needs to be emphasized here is that $\dii\mathcal{H}$
is identical to $\dii T_{00}$ and to the super-Hamiltonian $\dii H$
for the canonical scalar field~\eqref{canphi}, but differ for general
non-canonical fields.
[Note that $\dii\mathcal{H}$ is a Hamiltonian, while $\dii H$
is a Hamiltonian {\em density\/}.
Hence, the expressions~\eqref{eq:d2canH} and~\eqref{eq:d2SH}
differ by an overall factor of $a^3$.]
Also, the ratio of the factors in front of $\ddophi_k^2$ and $k^2\,\dophi_k^2$
is equal to $c_{\parallel}^2$ given in Eq.~\eqref{c2} and, thus, to the
speed of sound~\eqref{1stordersos} obtained from the equation of motion.
This implies that $\dii T_{00}$ and the canonical Hamiltonian $\dii\mathcal{H}$
become unstable under different conditions.
\par
The findings presented in this section are partly reminiscent of some
general results found by Maccallum and Taub~\cite{1972-Maccallum.Taub-CMP}
(see also Sec.~(13.2) in Ref.~\cite{1980-Kramer.etal-Exactsolutionsof}),
who showed that the variation of the action and the gauge fixing
(in this case, the symmetry reduction) do not necessarily commute,
hence the two procedures may not lead to the same equations of
motion.
To be precise, in Ref.~\cite{1972-Maccallum.Taub-CMP}, they found that
variation and reduction should be commuting operations for Class~A
space-times, to which the FRW~universe belongs,
but their result was obtained assuming
the presence of a standard fluid or a canonical scalar field.
In this sense, our results can be considered as an extension of their studies
to non-canonical Lagrangians, and show that the variation and gauge fixing
do not commute even in a FRW universe when the scalar field is not canonical.
\section{Discussion and examples}
Let us now come to the main points of discussion in this note.
The first question is why the canonical Hamiltonian, perturbed
stress~tensor and super-Hamiltonian coincide for a canonical scalar
field, but not for general scalar field Lagrangians.
To go about answering this question, it is necessary to look at the four
approaches we have employed from a different perspective.
In the first two approaches -- perturbed stress~tensor and covariant approach
-- we perturb the general expression for the stress~tensor of the
scalar field and obtain its second-order contribution  $\dii T_{00}$.
In the last two approaches -- ADM~formulation and symmetry-reduced action
-- we expand the action to second order in the
perturbation and obtain the super-(canonical) Hamiltonian of the
corresponding perturbed action.
While the super-Hamiltonian $\dii H$ is identical to $\dii T_{00}~$\footnote{In
Appendix~\ref{B}, we provide a general proof that the stress tensor and
super-Hamiltonian are equal to all orders in the FRW background.},
the canonical Hamiltonian $\dii\mathcal{H}$ is different.
\par
This then raises a related question:
Why is the super-Hamiltonian $\dii H$ different from the canonical
Hamiltonian $\dii\mathcal{H}$ for non-canonical scalar fields?
To answer this, let us assume that the coefficients containing the background
quantities $\PXbg$ and $\PXXbg$ in the second-order
action~(\ref{eq:action2order}) are constant and {\em independent\/} of $N$.
It is then easy to see that, using Eq.~(\ref{eq:def-superH}),
the resulting super-Hamiltonian is identical to the canonical
Hamiltonian~(\ref{eq:d2canH}) in this approximation.
In other words, the canonical Hamiltonian given in Eq.~(\ref{eq:d2canH})
is consistent provided the time-variation of background quantities
(like $\PXbg$ and $\PXXbg$) can be neglected.
(For the canonical scalar field, these functions are indeed constant and
the perturbed quantities therefore coincide.)
Although such an approximation may be valid for specific non-canonical
fields, they fail for some of the known fields used in the literature, as
we now proceed to review.
\subsection{$k$-essence}
\label{ksec}
Let us consider the simplest non-canonical scalar field discussed in
the context of power-law inflation,
\be
a(t) = a_0 \lt\frac{t}{t_0}\rt^{2/3\gamma}
\ ,
\qquad a_0 = a(t_0)
\ ,
\ee
whose Lagrangian
is~\cite{1999-Armendariz-Picon.etal-PLB}
\be
P = f(\phi) \left(X^2-X \right)
\ .
\ee
Upon solving the equation of state and the master equation
for the evolution of the energy density $\epsilon$
(as derived from the Einstein field equations),
\be
\epsilon + p = \gamma\, \epsilon
\ ,
\qquad
\dot\epsilon = -\sqrt{3\,\epsilon}\,\lp\, (\epsilon + p)
\ ,
\ee
one finds
\be
\Xbg = \frac{2-\gamma}{4-3\,\gamma}
\ ,
\label{Xexample}
\ee
so that
\beq
\dotp_0
\equiv
\sqrt{2\,X_0}
=
\sqrt\frac{4-2\,\gamma}{4-3\,\gamma}
\eeq
is constant and the background scalar field evolves in time according to
\beq
\pbg(t)
=
\sqrt{\frac{4-2\,\gamma}{4-3\,\gamma}}\, t
\ .
\eeq
One therefore finds that the background power-law ``potential''
also evolves in time, namely
\be
f(\pbg)
&\!\!=\!\!&
\frac{4\, (4-3\,\gamma)\, f_0}
{\left\{ 2\,\sqrt{4-3\,\gamma}
+ \sqrt{3\, f_0}\,\gamma\,\lp \lqu \pbg(t)-\pbg(t_0) \rqu\right\}^2}
\nonumber
\\
&\!\!=\!\!&
\frac{f_0}{\left[1+g_0\,(t-t_0)\right]^2}
\ ,
\ee
where $f_0 \equiv f(\pbg(t_0))$, and so evolve $\PXbg$ and $\PXXbg$. 
\par
In order to achieve an accelerated expansion,
\emph{i.e.}~inflation, $\gamma$ must range in $[0, 2/3]$
so that $\Xbg$ is inside the interval $[1/2, 2/3]$.
The explicit expressions for the various ``speeds of sound''
are [see Eqs.~\eqref{1stordersos}, \eqref{c1} and \eqref{crho},
respectively]
\be
&&
c^2_{\dophi}=\frac{2\,\Xbg-1}{6\,\Xbg-1}
\nonumber
\\
&&
c_{0}^2=-\frac{2\,\Xbg+1}{18\,\Xbg-1}
\\
&&
c_{\rho}^2 =-\frac{6\,\Xbg-1}{18\,\Xbg-1}
\ .
\nonumber
\ee
On substituting Eq.~\eqref{Xexample} and taking into account
the valid ranges for $\gamma$ and $\Xbg$ given above, one finds
\be
&
0<c^2_{\dophi} < \frac{1}{9}
&
\nonumber
\\
&
-\frac{1}{4}<c_{0}^2 < -\frac{7}{33}
&
\\
&
-\frac{3}{11} < c_{\rho}^2 < -\frac{1}{4}
\ .
&
\nonumber
\ee
Hence, $c_{0}$ is imaginary (equivalently, $\dii T_{00}$
and the super-Hamiltonian $\dii H$ are unstable) for all values of $\Xbg > 1/2$
allowed by the background dynamics, and, in particular, for those required
to achieve accelerated expansion.
The same occurs for $c_\rho$ (which is a decreasing function of $\Xbg$).
However, as is well known, the velocity $c_{\dophi}$ is real and
well-defined in the entire range of admissible $\gamma$ and $\Xbg$.
A possible physical interpretation of this finding is that, whereas
the field perturbation $\delta \phi$ propagates with real and
well defined velocity on the chosen background,
its energy density grows in time and drives the system out of
the perturbative regime.
This result cannot be just a curiosity and the instability must have
physical consequences,
given that gravity necessarily couples to the energy density.
\subsection{Tachyon}
For the tachyon~\cite{2002-Padmanabhan-PRD}, whose Lagrangian is
\be
P = - V(\phi)\,\sqrt{1 - 2\,X}
\ ,
\ee
where $V$ is positive in the background FRW,
the background dynamics requires that $X < 1/2$.
If one further imposes that $\dii T_{00}$ (and the super-Hamiltonian)
is stable (in the sense we already specified in the previous sections),
one obtains the new constraint $X<1/4$.
Here the requirement that the perturbations must be stable leads to a smaller
parameter range for the tachyonic field during inflation.
If one instead uses Eq.~\eqref{crho}, one finds that $X>1/2$ for $c_\rho$
to be positive, which is incompatible with the all of the values allowed by the
background dynamics mentioned above.
Again, $c_{\dophi}$ is instead always real.
\subsection{DBI field}
We also find a similar situation for the Dirac-Born-Infeld (DBI)
field~\cite{2004-Alishahiha.etal-PRD,2007-Chen.etal-JCAP},
whose Lagrangian is
\be
P = - \frac{1}{f(\phi)}\left(\sqrt{1 - 2 \, f(\phi) \, X} - 1\right) - V(\phi)
\ ,
\ee
with $f$ and $V$ positive functions in the background FRW
space-time.
The background dynamics requires that $X < 1/(2 f_{_0})$ but,
as above, if one further imposes that $\dii T_{00}$ (and the super-Hamiltonian)
be stable, one obtains the stronger constraint $X < 1/ (4 f_{_0})$.
Using Eq.~\eqref{crho}, one finds that $X>1/(2 f_{_0})$ for $c_\rho$
to be positive.
Just as it happens in the tachyonic case, this is incompatible with all
of the values allowed by the background dynamics.
Like in the previous examples, $c_{\dophi}$ is instead real and does
not introduce new constraints.
\section{Conclusions}
In this work, we have considered perturbations of a generalized scalar field
in four different approaches.
We have shown that second-order quantities obtained in these approaches
coincide for the canonical scalar field but are in general different.
At the root of the discrepancy lies the fact that, in evaluating the canonical
Hamiltonian from the second-order action, one implicitly assumes
that background quantities, like $\PXbg$ and $\PXXbg$, are constant.
As appears clearly, {\emph e.~g.}~for the $k$-essence reviewed in
Section~\ref{ksec}, background quantities are in general time-dependent
and neglecting this feature leads to incorrect expressions.
In particular, one expects the faster the background evolves, the
larger the discrepancy.
This aspect should be taken into account, for example, when
one applies adiabatic or slow-roll approximations.
\par
We have also shown that instabilities in general occur in the components
of the perturbed stress~tensor to second order, signaled by imaginary
{\em speeds of sound\/}~\eqref{eq:cs-def}.
Consequently, for specific known non-canonical Lagrangians,
demanding that the second-order stress~tensor be stable against small
field perturbations $\delta \phi$ restricts (and possibly rules out completely)
the parameter range of the scalar field.
Let us further recall that the energy density for the canonical scalar field
in the background frame is always stable, but instabilities appear in the
fluid-comoving frame, namely in the expressions for $\rho$ and $p$.
Our analysis indicates that, for instance, the results of
Refs.~\cite{2005-Seery.Lidsey-JCAP,2005-Gruzinov-PRD}
should be carefully reanalyzed in other approaches to check for plausible
inconsistencies.
Also, it is important to repeat the analysis
of Malik~\cite{2007-Malik-JCAP} -- by considering second-order
perturbations -- for general scalar fields.
\par
Our findings have been obtained by freezing the metric perturbations
completely and one could naturally wonder if the instabilities we found
would stand a more general investigation.
Unfortunately, a complete treatment of metric perturbations to second order 
is extremely involved and goes beyond the scope of the present work.
However, in Appendix~\ref{mp}, we provide some preliminary results
which suggest that metric perturbations should not affect the
aforementioned instabilities, because they do not seem to contribute
new terms to the ratios~\eqref{eq:cs-def}.
\par
Our results naturally raise the question of
{\it linearization instability\/} for non-canonical scalar
fields~\cite{1973-Fischer.Marsden-GRG}.
It has been known in the literature that the
linearization of non-linear fields can lead to spurious
solutions~\cite{1973-Brill.Deser-CMP,1973-Fischer.Marsden-GRG,1976-D'Eath-AP,
1986-Arms-APP}.
In particular, Brill and Deser~\cite{1973-Brill.Deser-CMP}
showed that there are spurious solutions to the linearized Einstein's equations
around the vacuum space-time given by the flat three-torus with zero extrinsic
curvature.
In the case of perturbations of the FRW background, D'Eath~\cite{1976-D'Eath-AP}
showed that the perturbations of Einstein's equations with the isentropic
perfect fluid matter is {\it linearization stable\/}.
(It is important to note that linearization instability is not directly related
to other kinds of instability.
However, dynamical instability is often studied by examining solutions of the
linearized equations~\cite{1986-Arms-APP}.)
In Einstein gravity with canonical fields, linearization instability
requires a compact space.
However the non-canonical scalar fields are non-linear, which
complicates the scenario.
Our analysis shows that the perturbation of these fields about the
spatially flat FRW background (by freezing the metric perturbations) might
still turn out to be {\it linearization unstable\/}.
It is thus important to repeat D'Eath's analysis for this class of scalar fields
by including the metric perturbations about the FRW~background.
We hope to address this issue further in a future publication.
\section*{Acknowledgement}
The authors gratefully acknowledge R.~Abramo, M.~Bruni,
F.~Finelli, L.~Hollenstein,
R.~Maartens, K.~Malik, G.~Marozzi, D.~Matravers, G.~Spada,
D.~Wands and R.~Woodard for comments on the manuscript
and useful discussions on the topic.
R.~C.~is supported by the INFN grant BO11
and S.~S.~by the Marie Curie Incoming International Grant
IIF-2006-039205.
\appendix
\section{Rescaled perturbation field}
\label{rescale}
Terms with $\ddophi$ can be eliminated from the equation of
motion~\eqref{1steom} by means of the rescaling~\eqref{chi} with
\be
r(t) = \exp \lt -\frac{1}{2} \int^t \frac{C}{A}\, \ud t' \rt
\ ,
\label{reom}
\ee
where $C$ was given in Eq.~\eqref{coeC} and $A=\PXbg  + \PXXbg\, \dpbg^2$
is the coefficient of $\dddophi$ in Eq.~\eqref{1steom}.
The equation for the rescaled field thus reads
\begin{subequations}
\be
A\,\ddot\chi+\frac{B}{a^2}\,\chi_{,ii}+\tilde D\,\chi=0
\ ,
\ee
where $B=-\PXbg$ and
\be
\tilde D=D+\frac{C\,\dot A}{2\,A}
-\frac{C^2}{4\,A}-\frac{\dot C}{2}
\ .
\ee
\end{subequations}
This operation therefore modifies the coefficient of $\dophi$ (the ``mass term''),
leaving the speed of sound in the equation for $\chi$ the same as the one
given in Eq.~\eqref{1stordersos}.
\par
By similar treatments, terms linear in $\ddophi$ can be eliminated from the
components of the stress tensor.
For example, $\dii T_{00}$ has the following structure
\be
\dii T_{00} =
\bar A\,\ddophi^2 + \frac{\bar B}{a^2}\,\dophi_{,i}^2 +\bar C\,\dophi\,\ddophi
+\bar  D\,\dophi^2
\ ,
\ee
where the time-dependent coefficients $\bar A$, $\bar B$, $\bar C$ and
$\bar D$ can be read from Eq.~\eqref{eq:d2T00}.
The rescaling in Eq.~\eqref{chi}, now with
\be
r(t) = \exp \lt -\frac{1}{2} \int^t \frac{\bar C}{\bar A} \, \ud t' \rt
\ ,
\label{rT}
\ee
then yields
\be
\dii T_{00} = \bar A\, \dot\chi^2 + \frac{\bar B}{a^2}\,\chi_{,i}^2
+\lt \bar D - \frac{\bar C^2}{4\,\bar A} \rt \chi^2
\ ,
\ee
or, in the Fourier domain,
\be
\dii T_{00} = \bar A\,\dot\chi_k^2
+ \lt \bar B\,\frac{k^2}{a^2} +
\bar D - \frac{\bar C^2}{4\,\bar A} \rt \chi_k^2
\ .
\ee
Hence, $\dii T_{00}$ is stable when
\be
\frac{k^2}{a^2}
\ge
\frac{\bar C^2}{4\,\bar A\,\bar B}-\frac{\bar D}{\bar B}
\ .
\label{condstab}
\ee
Eq.~\eqref{condstab} can be regarded as a test that $P$
(and its derivatives) must pass if one wants to deal with
a model endowed with a stable perturbation theory.
Similar conditions can be found for other quantities,
like $\dii \rho$ in Eq.~\eqref{eq:delta2rho} or $\dii p$
in Eq.~\eqref{eq:delta2p}.
Let us also note in passing that, when the inequality~\eqref{condstab}
is saturated, the perturbation $\chi_k$ appears in a state of
``asymptotic freedom'' of the sort discussed in Ref.~\cite{2008-Appignani}.
\par
To summarise, although the transformation~\eqref{chi} with
$r$ given in Eq.~\eqref{reom} or Eq.~\eqref{rT} changes the values of
the stress tensor, density and pressure, it does not affect the
expressions of the corresponding velocities
and cannot remove the instabilities signalled by imaginary
speeds of  propagation.
\section{Equivalence between $T_{00}$ and $\hub$}
\label{B}
The equivalence between $T_{00}$ and the (matter part of the) super-Hamiltonian
$\hub$ can be proven in general in an FRW space-time in the proper time gauge.
From the standard definition for $T_{00}$,
\be
T_{00} = \frac{2}{\sqrt{-g}}\,\frac{\delta S}{\delta g^{00}}
\ ,
\ee
we find, for the metric \eqref{eq:FRW},
\be
T_{00} = - \frac{N^2}{a^3}\,\frac{\delta S}{\delta N}
\equiv \frac{N^2}{a^3}\,H
\ .
\ee
The energy contained in a given spatial (comoving) volume $V$ is equal to
\be
E = \int_V \ud^3 x\,\sqrt{-g} \, T_{00}
= \int_V \ud^3 x\,N^3\,H
= V\,N^3\,H
\ ,
\ee
where $V$ can be set equal to one without loss of generality.
It follows that, in the proper time gauge ($N=1$), one has $E=H$,
namely the super-Hamiltonian always equals the energy whose
density is given by $T_{00}$.
\section{Metric perturbations: preliminary results}
\label{mp}
A complete description of metric perturbations to second
order remains outside the scope of the present work and
we just outline the general features one might
encounter by including the metric perturbations
in two approaches: the perturbed stress~tensor and reduced action. 
\par
For the general Lagrangian~\eqref{eq:genscal}, the stress~tensor
is given by Eq.~\eqref{eq:Tmunu}.
The general perturbation about the FRW~background, {\em i.~e.},
\be
\begin{array}{l}
\phi = \phi_{_0}(t) + \epsilon \, \delta \phi(t,{\bf x})
\\
\\
g_{\mu\nu} = g_{\mu\nu}^{(0)}(t) + \epsilon \, \delta g_{\mu\nu}(t,{\bf x})
\ ,
\end{array}
\label{eq:metpert}
\ee	
leads to 
\begin{eqnarray}
T_{_{\mu\nu}}
&\!\!=\!\!&
\left.T_{\mu\nu}\right|_0
+ \epsilon\left[
\left.\frac{\partial T_{\mu\nu}}{\partial X} \right|_{0} 
\!\!\delta X
+\left. \frac{\partial T_{\mu\nu}}{\partial \phi} \right|_{0}\!\! \delta \phi
+\left.\frac{\partial T_{\mu\nu}}{\partial g_{\a\b}} \right|_{0} 
\!\! \delta g_{\a\b}
\right] 
\nonumber
\\
&&
+\frac{\epsilon^2}{2}\left[
\left. \frac{\partial^2 T_{\mu\nu}}{\partial X^2} \right|_{0} \!\! (\delta X)^2 
+ \left.\frac{\partial^2 T_{\mu\nu}}{\partial \phi^2} \right|_{0}\!\!  (\delta \phi)^2
\right.
\nonumber
\\
&&
\phantom{+\frac{\epsilon^2}{2}}\ 
\left.
+\left.\frac{\partial^2 T_{\mu\nu}}{\partial g_{\a\b}^2}\right|_{0} 
\!\!(\delta g_{\a\b})^2
\right]
+ {\rm~cross~terms}
\ ,
\end{eqnarray}
where $|_0$ means the expression is evaluated on the background
quantities $\phi_0$, $X_0$ and $g_{\a\b}^{(0)}$.
Focusing on the metric perturbations, we have
\be
\frac{\partial T_{\mu\nu}}{\partial g_{\a\b}}
&\!\!=\!\!&
\frac{\partial^{\a}\phi\, \partial^{\b}\phi}{2}
\left( P_{_{XX}}\, \partial_{\mu}\phi \, \partial_{\nu}\phi- g_{\mu\nu}\, P_{_{X}} \right) 
- \delta^{\a}_{\mu} \, \delta^{\b}_{\nu}\, P
\nonumber
\\
\frac{\partial^2 T_{\mu\nu}}{\partial g_{\gamma \delta}\, \partial g_{\a\b}}
&\!\!=\!\!& 
\frac{\partial^{\a}\phi\, \partial^{\b}\phi\, \partial^{\gamma}\phi\, \partial^{\delta}\phi}{4}
\\
&&
\times  
\left(P_{_{XXX}}\, \partial_{\mu}\phi \, \partial_{\nu}\phi - g_{\mu\nu}\,P_{_{XX}} \right) 
\nonumber
\\
& & - \frac{P_{_X}}{2} \left(\delta^{\a}_{\mu}\, \delta^{\b}_{\nu} \, 
\partial^{\gamma}\phi\, \partial^{\delta}\phi + \delta^{\gamma}_{\mu}\, \delta^{\delta}_{\nu}
\, \partial^{\a}\phi\, \partial^{\b}\phi
\right)
\ .
\nonumber
\ee
Evaluating the above expressions in the FRW background, we get
\be
\left. \frac{\partial T_{\mu\nu}}{\partial g_{\a\b}} \right|_{_{\rm FRW}} 
\!\!\!\!\!\!&\!\!=\!\!&
\frac{\delta^{\a}_{0}\, \delta^{\b}_{0}\, \dpbg^2}{2}
\left(\PXXbg\, \dpbg^2 \delta^{0}_{\mu}\,\delta^{0}_{\nu}
-g_{\mu\nu}^{(0)}\, \PXbg  \right)
\nonumber
\\
&&
-\delta^{\a}_{\mu}\,\delta^{\b}_{\nu}\,\Pbg
\nonumber
\\
\left. \frac{\partial^2 T_{\mu\nu}}{\partial g_{\gamma \delta}\,\partial g_{\a\b}}
\right|_{_{\rm FRW}} 
\!\!\!\!&\!\!=\!\!&
\frac{\delta^{\a}_{0}\, \delta^{\b}_{0}\,\delta^{\gamma}_{0}\, \delta^{\delta}_{0}\,\dpbg^4}{4}
\\
&&
\times  
\left( \PXXXbg\,\dpbg^2\, \delta_{\mu}^0\, \delta_{\nu}^0
- g_{\mu\nu}^{(0)}\, \PXXbg \right)
\nonumber
\\
&&
-\frac{\PXbg}{2} \left(
\delta^{\a}_{\mu}\, \delta^{\b}_{\nu}\, \delta^{\gamma}_{0}\, \delta^{\delta}_{0}
+ \delta^{\gamma}_{\mu}\, \delta^{\delta}_{\nu} \,  \delta^{\a}_{0}\, \delta^{\b}_{0}
\right)
\dpbg^2
\ .
\nonumber
\ee
For the $00$-component of the stress~tensor, we then have
\begin{eqnarray}
\left. \frac{\partial T_{00}}{\partial g_{\a\b}} \right|_{_{\rm FRW}}
\!\!\!\!
&\!\!=\!\!&
\frac{\delta^{\a}_{0}\, \delta^{\b}_{0}\, \dpbg^2}{2}
\left(\PXXbg\, \dpbg^2 - a^2 \, \PXbg  \right)  
- \delta^{\a}_{0} \, \delta^{\b}_{0}\, \Pbg
\nonumber
\\
\left. \frac{\partial^2 T_{00}}{\partial g_{\gamma \delta}\, \partial g_{\a\b}}
\right|_{_{\rm FRW}} 
\!\!\!\! &\!\!=\!\!&
\frac{\delta^{\a}_{0}\, \delta^{\b}_{0}\, \delta^{\gamma}_{0}\, \delta^{\delta}_{0} }{4}
\label{eq:d2T00g}
\\
&&
\times  
\left( \PXXXbg\, \dpbg^6 - a^2\, \PXXbg\, \dpbg^4 - 4\, \PXbg \, \dpbg^2 \right)
\ .
\nonumber
\ee
By comparing the second-order expression of the
stress~tensor~\eqref{eq:d2T00} with Eq.~\eqref{eq:d2T00g},
we immediately see a peculiar difference:
$\PXXXbg$ in Eq.~\eqref{eq:d2T00g} multiplies $\dpbg^6$,
whereas it multiplies $\dpbg^4$ in Eq.~\eqref{eq:d2T00}.
This means that metric perturbations and matter perturbations
give rise to different contributions to the stress~tensor and
including the metric perturbations should not, in principle, remove
the instability we found at second order. 
Moreover, Eq.~\eqref{eq:d2T00g} makes it clear that investigating
the complete second-order perturbations is highly non-trivial.
\par
We finally look at the metric perturbations in the reduced action 
approach.
On substiuting~\eqref{eq:metpert} in Eq.~\eqref{eq:genscal},
the Lagrangian can be written as
\be
P
&\!\!=\!\!&
\Pbg
+ \epsilon\left[\left.\frac{\partial P}{\partial X} \right|_{0}\!\!\delta X
+\left. \frac{\partial P}{\partial \phi} \right|_{0}
\!\! \delta \phi 
+\left.\frac{\partial P}{\partial g_{\mu\nu}} \right|_{0} 
\!\! \delta g_{\mu\nu}
\right]
\nonumber
\\
&&
+\frac{\epsilon^2}{2}\left[
\left.\frac{\partial^2 P}{\partial X^2} \right|_{0} 
\!\! (\delta X)^2 
+\left.\frac{\partial^2 P}{\partial \phi^2} \right|_{0} 
\!\!(\delta \phi)^2
\right.
\\
&&
\phantom{+\frac{\epsilon^2}{2}}\
\left.
+\left.\frac{\partial^2 P}{\partial g_{\gamma \delta}\, \partial g_{\a\b}}\right|_{0} 
\!\! \delta g_{\a\b} \, \delta g_{\gamma \delta}
\right]
+ {\rm~cross~terms}
\ .
\nonumber
\end{eqnarray}
Again, focusing on the metric perturbations, we have
\be
\begin{array}{c}
\strut\displaystyle\frac{\partial P}{\partial g_{\a\b}}
=
\frac{P_{_{X}}}{2} \, \partial^{\a} \phi\, \partial^{\b} \phi
\\
\\
\strut\displaystyle
\frac{\partial^2 P}{\partial g_{\alpha\beta}\, \partial g_{\gamma \delta}}  
=
\frac{1}{4}\, \partial^{\alpha} \phi\, \partial^{\beta} \phi \,
\partial^{\gamma} \phi\, \partial^{\delta} \phi \,P_{_{XX}} 
\ . 
\end{array}
\ee
The above expressions in the FRW background yield
\be
\begin{array}{c}
\strut\displaystyle
\left. \frac{\partial P}{\partial g_{\mu\nu}} \right|_{_{\rm FRW}}
\!\!\!\!= 
\frac{P_{_{X}}}{2} \, \delta^{\mu}_0\, \delta^{\nu}_0\, \dpbg^2 
\\
\\
\strut\displaystyle
\left. \frac{\partial^2 P}{\partial g_{\alpha\beta}\, \partial g_{\gamma \delta}}
\right|_{_{\rm FRW}} 
\!\!\!\!= 
\frac{1}{4}\, \delta^{\alpha}_0\, \delta^{\beta}_0\,
\delta^{\gamma}_0 \delta^{\delta}_0  \,  \dpbg^4 \, P_{_{XX}} 
\ .
\end{array}
\label{eq:d2Pg}
\ee
Comparing the above expressions with Eq.~\eqref{eq:d2p}, we note that
the contributions of the metric perturbations to the second-order action
differ from the contributions of the metric perturbations to the second-order
stress~tensor.
\par
From the above preliminary analysis,  it seems that the features we
have obtained in this work continue to hold with the metric perturbations.
However, this issue needs a thorough investigation and will be 
discussed elsewhere.

\end{document}